\documentclass[pre,twocolumn,floats,superscriptaddress]{revtex4}
\usepackage{amsmath}
\usepackage{amssymb}
\usepackage{graphicx}
\usepackage{color}

\begin{document}

\title{Deterministic dynamics of overactive Brownian particle in 2D and 3D potential wells}
\author{Denis S.\ Goldobin}
\affiliation{Institute of Continuous Media Mechanics, UB RAS, Academician Korolev
Street 1, 614013 Perm, Russia}
\affiliation{Department of Control Theory, Nizhny Novgorod State University, Gagarin Avenue 23, 603022 Nizhny Novgorod, Russia}
\affiliation{Department of Theoretical Physics, Perm State University, Bukirev
Street 15, 614990 Perm, Russia}
\email{denis.goldobin@gmail.com}
\author{Lev A.\ Smirnov}
\affiliation{Department of Control Theory, Nizhny Novgorod State University, Gagarin Avenue 23, 603022 Nizhny Novgorod, Russia}
\author{Lyudmila S.\ Klimenko}
\affiliation{Institute of Continuous Media Mechanics, UB RAS, Academician Korolev
Street 1, 614013 Perm, Russia}
\affiliation{Department of Theoretical Physics, Perm State University, Bukirev
Street 15, 614990 Perm, Russia}
\author{Grigory V.\ Osipov}
\affiliation{Department of Control Theory, Nizhny Novgorod State University, Gagarin Avenue 23, 603022 Nizhny Novgorod, Russia}
\date{\today}

\begin{abstract}
We study deterministic dynamics of overactive Brownian particles in 2D and 3D potentials. This dynamics is Hamiltonian. Integrals of motion for continuous rotational symmetries are reported. The cases of 2D, axisymmetric and non-axisymmetric 3D potentials are characterized and compared to each other. The strong impact of the rotational symmetry integrals of motion on the chaotic and quasiperiodic orbits is revealed. The scattering cross section is reported for spherically symmetric Gaussian-shape potential wells and obstacles as a function of self-propulsion speed.
\end{abstract}

\maketitle

For diverse problems of motion of self-propelled particles in a heterogeneous medium or on a disordered substrate, the problem of the particle scattering on an isolated impurity or defect is relevant as a fundamental theoretical ingredient~\cite{heter_med1,heter_med2,heter_med3,heter_med4,heter_med5,heter_med6}. This problem turns out to be surprisingly versatile and sophisticated even for the most basic case, where the thermal fluctuations are neglected and purely deterministic dynamics is considered. The latter is a physically reasonable approximation for overactive Brownian particles, the size of which makes the effect of thermal fluctuations not so much overwhelming, while the locomotion forces create a considerable propulsion speed.

In this paper we consider the deterministic dynamics of an overactive Brownian particle in a 2D potential well, in axisymmetric and non-axisymmetric 3D potential wells; the problem of scattering on spherically symmetric potential wells and obstacles.

\section{Deterministic dynamics of active Brownian particles}
In this section we consider a specific physical setup which can lead to the mathematical reduction we consider in the following sections. This derivation from first principles is needed for the cases where the reduction require amendments, accounting for thermal noise, {\it etc.}

For micromotors---particles of a micrometer size which are able to move progressively through the fluid without external forcing or superimposed gradient of the medium parameters---several self-propulsion mechanism are possible. However, the promising technical application (drug delivery~\cite{drug-delivery1,drug-delivery2}, smart materials~\cite{smart-materials}, {\it etc.}) are related to the artificial microparticles, where the life-like mechanisms (squirming motion~\cite{squirming} or ciliary propulsion~\cite{ciliary}) are less feasible than the catalytic motors~\cite{catalyt_mot1,catalyt_mot2,catalyt_mot3,catalyt_mot4,catalyt_mot5}.

The standard design for catalytic motors (Janus- particles and droplets) is a heterogeneous catalytically active surface inducing a matter flow in a reactive surrounding liquid.
\begin{description}
  \item[(i)]
On the micrometer scale, the induced flow is a quasi-stationary Stokes viscous flow, which generates a drug force with a nearly constant absolute value the direction of which is frozen into the particle and subject to change along with the change of the particle orientation.
  \item[(ii)]
The manufactured particles with a heterogeneous surface are not spherically symmetric, but possess an axial symmetry, which is natural for the feasible designs of mass production of microobjects. The ambient liquid flow exerts a Stokes viscous torque on the particle, when the particle symmetry axis deviates from the propulsion direction. Generally, depending on the particle shape, this torque can make the particle oriented against the incident flow, along it, or transversal to it. For technical applications only the former case is potentially interesting. Therefore, one can restrict consideration to the case, where the particle tends to orient itself along the propulsion direction ({\em i.e.}, against the incident flow).
\end{description}
The physical setup (i) and (ii)---an axisymmetric catalytic particle experiencing a constant drug force $f_\mathrm{cd}$ along its director $\mathbf{n}$, the Stokes viscous friction force, and Stokes viscous friction torque orienting the director along the propulsion direction---is governed by the following mathematical model:
\begin{align}
\mu\frac{\mathrm{d}\mathbf{v}}{\mathrm{d}t}&=f_\mathrm{cd}\mathbf{n}-\gamma\mathbf{v}+\mathbf{f}\,,
\label{eq:ph:01}\\
J_\perp\frac{\mathrm{d}^2\mathbf{n}}{\mathrm{d}t^2}&=
-\Gamma_\perp\frac{\mathrm{d}\mathbf{n}}{\mathrm{d}t} +K_\perp\big(\mathbf{v}-\mathbf{n}(\mathbf{n}\cdot\mathbf{v})\big)\,,
\label{eq:ph:02}
\end{align}
where $\mu$ is the particle mass, $\mathbf{f}$ is the external force, $J_\perp$ is the moment of inertia about the axis perpendicular to the symmetry axis, $\Gamma_\perp$ is the Stokes friction torque coefficient for the rotation about the same axis, the rotation about the symmetry axis (about $\mathbf{n}$) only decays and is omitted.
\begin{itemize}
\item
In~(\ref{eq:ph:01}) for an axisymmetric particle, in place of $\gamma\mathbf{v}$, one should generally write the Stokes force $\mathbf{\gamma}\cdot\mathbf{v}$ with $\mathbf{\gamma}=\gamma_\parallel\mathbf{n}\mathbf{n}+\gamma_\perp(\mathbf{I}-\mathbf{n}\mathbf{n})$, where $\mathbf{I}$ is the identity matrix, but the mismatch $\gamma_\perp\ne\gamma_\parallel$ does not make dramatic changes to the dynamics in the case of $\mathbf{v}$ weakly deviating from $\mathbf{n}$. Hence, for the sake of simplicity, we derive equations for $\gamma_\perp=\gamma_\parallel=\gamma$.
\item
Further, we explain the last term of Eq.~(\ref{eq:ph:02}). The viscous Stokes flow generally creates a linear in $\mathbf{v}$ torque $\mathbf{K}\cdot\mathbf{v}$; for the axisymmetric particle, the coefficient matrix simplifies to $\mathbf{K}=K_\parallel\mathbf{n}\mathbf{n}+K_\perp(\mathbf{I}-\mathbf{n}\mathbf{n})$. However, for the particles we consider, this torque disappears as long as the particle is oriented along the propulsion direction, $\mathbf{v}=v\mathbf{n}$, where one can calculate $\mathbf{K}\cdot v\mathbf{n}=K_\parallel v\mathbf{n}$. Hence, the coefficient $K_\parallel=0$ and the term $\mathbf{K}\cdot\mathbf{v}=K_\perp\big(\mathbf{v}-\mathbf{n}(\mathbf{n}\cdot\mathbf{v})\big)$.
\item
In Eq.~(\ref{eq:ph:01}), there are no terms proportional to $\mathrm{d}\mathbf{n}/\mathrm{d}t$; no integral Stokes force emerges for rotation of, at least, the ellipsoidal particles or, more generally, the axisymmetric particles with additional mirror-symmetry with respect to their equator plane.
\end{itemize}

In the overdamped ({\it or} small inertia) limit, which is relevant for the Brownian particles, one neglects the $\mu$- and $J_\perp$-terms [as $\mu/(\gamma\tau)\ll1$ and $J_\perp/(\Gamma_\perp\tau)\ll1$, where $\tau$ is the dynamics reference time scale] and finds:
\begin{align}
\frac{\mathrm{d}\mathbf{x}}{\mathrm{d}t}=\mathbf{v}& =\frac{f_\mathrm{cd}\mathbf{n}+\mathbf{f}}{\gamma}\,,
\label{eq:ph:03}\\
\frac{\mathrm{d}\mathbf{n}}{\mathrm{d}t}& =\frac{K_\perp}{\Gamma_\perp}\big(\mathbf{v}-\mathbf{n}(\mathbf{n}\cdot\mathbf{v})\big)
\nonumber\\
&=\frac{K_\perp}{\Gamma_\perp}\frac{\mathbf{f}-\mathbf{n}(\mathbf{n}\cdot\mathbf{f})}{\gamma}\,.
\label{eq:ph:04}
\end{align}
With $\mathbf{v}=V\mathbf{e}$, $|\mathbf{e}|=1$, Eq.~(\ref{eq:ph:03}) can be recast as
\begin{equation}
V\mathbf{e}=\frac{f_\mathrm{cd}}{\gamma}\mathbf{n}+\frac{\mathbf{f}}{\gamma}\,.
\label{eq:ph:05}
\end{equation}
In the limiting case of ``overactive'' particle, $f\ll f_\mathrm{cd}$,
$V^2=(f_\mathrm{cd}^2+2f_\mathrm{cd}(\mathbf{n}\cdot\mathbf{f}) +f^2)/\gamma^2$ and
\[
V=\frac{f_\mathrm{cd}+(\mathbf{n}\cdot\mathbf{f}) +\mathcal{O}(f^2/f_\mathrm{cd})}{\gamma}\,.
\]
Hence,
\begin{align}
\mathbf{e}&=\frac{f_\mathrm{cd}\mathbf{n}+\mathbf{f}}
{f_\mathrm{cd}+(\mathbf{n}\cdot\mathbf{f}) +\mathcal{O}(f^2/f_\mathrm{cd})}
\nonumber\\
&=\mathbf{n}+\frac{\mathbf{f}-\mathbf{n}(\mathbf{n}\cdot\mathbf{f})}{f_\mathrm{cd}} +\mathcal{O}_1\left(\frac{f^2}{f_\mathrm{cd}^2}\right)\,,
\label{eq:ph:06}\\
\mbox{or }\quad
\mathbf{n}&
 =\mathbf{e}-\frac{\mathbf{f}-\mathbf{e}(\mathbf{e}\cdot\mathbf{f})}{f_\mathrm{cd}} +\mathcal{O}_2\left(\frac{f^2}{f_\mathrm{cd}^2}\right)\,.
\label{eq:ph:07}
\end{align}
The time-derivative of (\ref{eq:ph:06}) with substitution of (\ref{eq:ph:04}) and (\ref{eq:ph:07}) yields:
\begin{align}
\frac{\mathrm{d}\mathbf{e}}{\mathrm{d}t}
&=\frac{\mathrm{d}\mathbf{n}}{\mathrm{d}t}
+\frac{\frac{\mathrm{d}\mathbf{f}}{\mathrm{d}t} -\mathbf{n}(\mathbf{n}\cdot\frac{\mathrm{d}\mathbf{f}}{\mathrm{d}t})
-\frac{\mathrm{d}\mathbf{n}}{\mathrm{d}t}(\mathbf{n}\cdot\mathbf{f})
-\mathbf{n}(\frac{\mathrm{d}\mathbf{n}}{\mathrm{d}t}\cdot\mathbf{f})}{f_\mathrm{cd}}
+\dots
\nonumber\\
&=\frac{K_\perp}{\Gamma_\perp}\frac{\mathbf{f}-\mathbf{n}(\mathbf{n}\cdot\mathbf{f})}{\gamma}
+\frac{\mathbf{I}-\mathbf{n}\mathbf{n}}{f_\mathrm{cd}}\cdot(\mathbf{v}\cdot\nabla)\mathbf{f}
+\mathcal{O}\left(\frac{f^2}{f_\mathrm{cd}^2}\right)
\nonumber\\
&=\frac{K_\perp}{\Gamma_\perp}\frac{\mathbf{f}-\mathbf{e}(\mathbf{e}\cdot\mathbf{f})}{\gamma}
+\mathcal{O}\left(\frac{f^2}{f_\mathrm{cd}^2}\right)\,;
\label{eq:ph:08}
\end{align}
Eq.~(\ref{eq:ph:03}) reads
\begin{align}
\frac{\mathrm{d}\mathbf{x}}{\mathrm{d}t}&=V\mathbf{e}\,,
\label{eq:ph:09}
\\
V&=\frac{f_\mathrm{cd}+(\mathbf{e}\cdot\mathbf{f}) +\mathcal{O}(f^2/f_\mathrm{cd})}{\gamma}\,.
\label{eq:ph:10}
\end{align}

Eqs.~(\ref{eq:ph:08})--(\ref{eq:ph:10}) constitute the approximate model reduction for the deterministic dynamics of an overactive Brownian particle (catalytic motor, OAP) subject to external force $\mathbf{f}$.

The provided consideration allows one to deal with the cases where the overactive reduction is physically insufficient (as for the strait line trajectories in~\cite{Pikovsky-2023}) and introduce the thermal noise in accordance with the Fluctuation--Dissipation Theorem.

\section{Overactive limit: Hamiltonian dynamics}
At the limit $f/f_\mathrm{cd}\to0$, after rescaling, the dynamical system~(\ref{eq:ph:08})--(\ref{eq:ph:10}) can be recast as follows (for the ease of comparison to~\cite{Aranson-Pikovsky-2022}):
\begin{align}
  \dot{x}_j &=V n_j\;,
\label{eq001}
\\
  \dot{n}_j &=f_j-(\mathbf{f}\cdot\mathbf{n})n_j\;,
\label{eq002}
\\
U &=-U_0e^{-\frac{\sum_j\beta_jx_j^2}{2}}\;,
\label{eq003}
\\
f_j&=-\frac{\partial U}{\partial x_j}
\label{eq004}
\\
&=\beta_jx_jU\;,
%\label{eq0041}
\nonumber
\end{align}
where $V$ is a constant speed of particle self-propulsion, $U_0=(K_\perp u_0)/(\Gamma_\perp\gamma)$, and $u_0$ is the dimensional depth of the potential well. We will devote much attention to a specific class of Gaussian-shape potential wells~(\ref{eq003}) which are typically suggested in the literature for the laser-made optical traps in experiments with Brownian particles.

With rescalling $(x\to Lx,t\to \tau t)$, where $\beta_1L^2=1$ and $\tau=L/V$, one finds
\begin{align}
  \dot{x}_j &= n_j\;,
\label{eq005}
\\
  \dot{n}_j &=f_j-(\mathbf{f}\cdot\mathbf{n})n_j\;,
\label{eq006}
\\
U &=-\frac{U_0}{V}e^{-\frac{\sum_j\beta_jL^2x_j^2}{2}}\;.
\label{eq007}
\end{align}
The dimensionless problem is controlled by $(U_0/V)$ and $\beta_{2,3}/\beta_1$.

If the potential well is deep or $V$ is small, $U_0/V\gg1$, one finds $U=\frac{U_0}{V}\frac{\sum_j\beta_jL^2x_j^2}{2}$, which suggests an alternative choice of $L^2=V/(\beta_1U_0)$. With this choice, the limit $V/U_0\to 0$ provides a scaling invariant picture of dynamics: Eqs.~(\ref{eq001},\ref{eq002},\ref{eq004}) with $V=1$, $\beta_1=1$, \begin{equation}
U=\frac{x^2+\beta_2y^2+\beta_3z^2}{2}\,.
\label{eq:Ulim}
\end{equation}

Dynamical system~(\ref{eq001},\ref{eq002}) is Hamiltonian;
\begin{align}
\dot{x}_j&=\frac{\partial H}{\partial p_j}\,,
\qquad
\dot{p}_j=-\frac{\partial H}{\partial x_j}\,,
\label{eq008}
\\
&\mathbf{n}=\frac{\mathbf{p}}{|\mathbf{p}|}\,,
\label{eq009}
\\
H&=V|\mathbf{p}|-\exp\left(-\frac{U(\mathbf{x})}{V}\right)=0\,.
\label{eq010}
\end{align}
Indeed, $\dot{\mathbf{n}}=\frac{\dot{\mathbf{p}}}{|\mathbf{p}|} -\frac{\mathbf{p}(\mathbf{p}\cdot\dot{\mathbf{p}})}{|\mathbf{p}|^3} =\frac{\dot{\mathbf{p}}-\mathbf{n}(\mathbf{n}\cdot\dot{\mathbf{p}})}{|\mathbf{p}|} =-\frac{1}{|\mathbf{p}|V}e^{-U/V}(\mathbf{I}-\mathbf{n}\mathbf{n})\cdot\frac{\partial U}{\partial\mathbf{x}}\overset{H=0}{=}\mathbf{f}-(\mathbf{f}\cdot\mathbf{n})\mathbf{n}$.

Note, the dynamical system~(\ref{eq001},\ref{eq002},\ref{eq004}) is not equivalent to a general Hamiltonian system~(\ref{eq008}) with (\ref{eq009}) and given Hamiltonian function $H(\mathbf{x},\mathbf{p})$; it is only the flow of the latter Hamiltonian system on an invariant co-dimension $1$ manifold of (\ref{eq008}) with $H(\mathbf{x},\mathbf{p})=0$. Exactly the same mathematical formalism one can also find in geometrical optics with heterogeneity of the refractive index in place of $U(\mathbf{x})$~\cite{geom-opt}.

\subsection{Additional integrals of motion}
For a Hamiltonian dynamics, continuous shift symmetries introduce additional integrals of motion. Namely:
\\
$\bullet$\ In the case of {\em spherically symmetric} potential $U(|\mathbf{x}|)$, there must be rotational symmetries and associate integrals of motion:
\begin{equation}
\mathbf{M}=[\mathbf{x}\times\mathbf{p}]\,.
\label{eq:CS:1}
\end{equation}
Indeed, $\dot{\mathbf{M}}=[\dot{\mathbf{x}}\times\mathbf{p}]+[\mathbf{x}\times\dot{\mathbf{p}}] =[\mathbf{n}\times\mathbf{p}] -[\mathbf{x}\times\frac{\partial H}{\partial|\mathbf{x}|}\frac{\mathbf{x}}{|\mathbf{x}|}]=0$.
\\
$\bullet$\ In the case of {\em axisymmetric} potential $U(r_{xz},y)$, there is one rotational symmetry and an associate integral of motion:
\begin{align}
M_y&=(\mathbf{e}_y\cdot[\mathbf{x}\times\mathbf{p}])
\label{eq:AS:1}
\\
&=(\mathbf{e}_y\mathbf{x}\mathbf{p})
\,,
\nonumber
\end{align}
where $\mathbf{e}_y$ is the unit vector along the $y$-axis.
Indeed, $\dot{M}_y=(\mathbf{e}_y\dot{\mathbf{x}}\mathbf{p})+(\mathbf{e}_y\mathbf{x}\dot{\mathbf{p}}) =(\mathbf{e}_y\mathbf{n}\mathbf{p}) -\big(\mathbf{e}_y(\mathbf{r}_{xz}+y\mathbf{e}_y)\frac{\partial H}{\partial r_{xz}}\frac{\mathbf{r}_{xz}}{r_{xz}}\big) -(\mathbf{e}_y\mathbf{x}\frac{\partial H}{\partial y}\mathbf{e}_y)=0$, where $\mathbf{x}=\mathbf{r}_{xz}+y\mathbf{e}_y$.

\subsection{2D potential}
\begin{figure}[!htb]
{\sf (a)}\hspace{-15pt}
\includegraphics[width=0.47\textwidth]{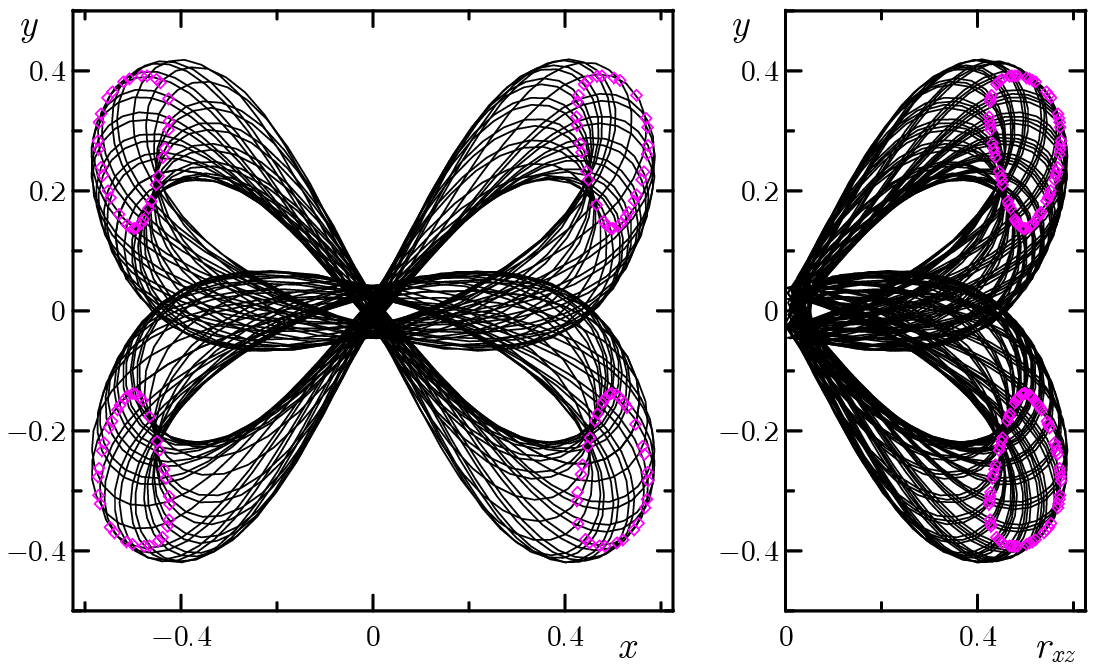}

\vspace{5pt}
{\sf (b)}\hspace{-15pt}
\includegraphics[width=0.47\textwidth]{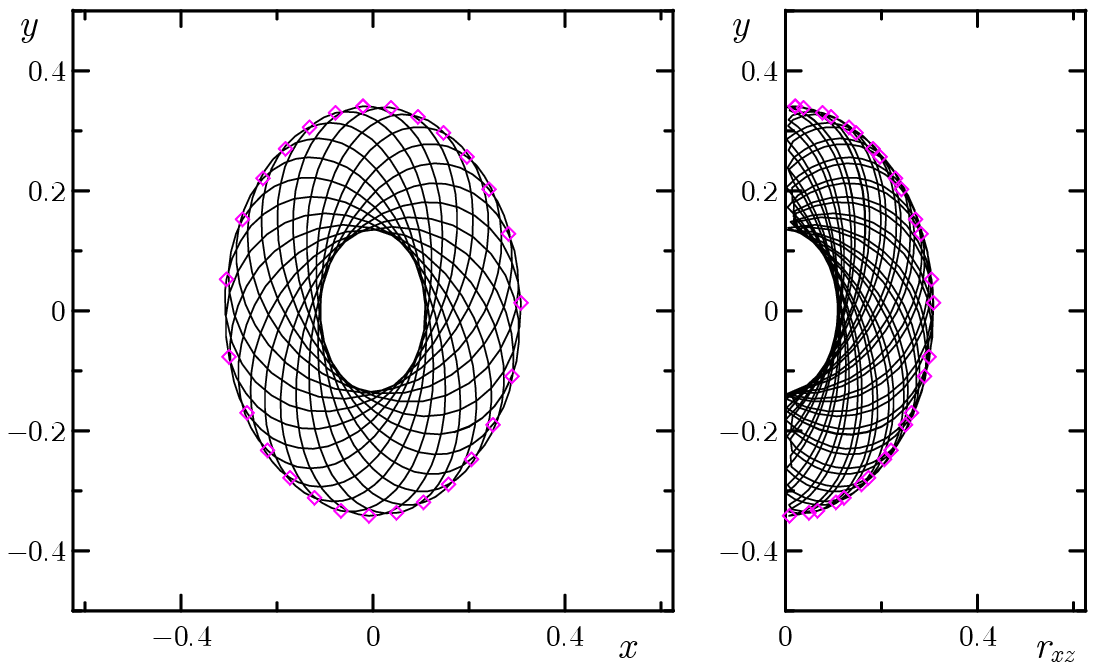}

\vspace{5pt}
{\sf (c)}\hspace{-15pt}
\includegraphics[width=0.47\textwidth]{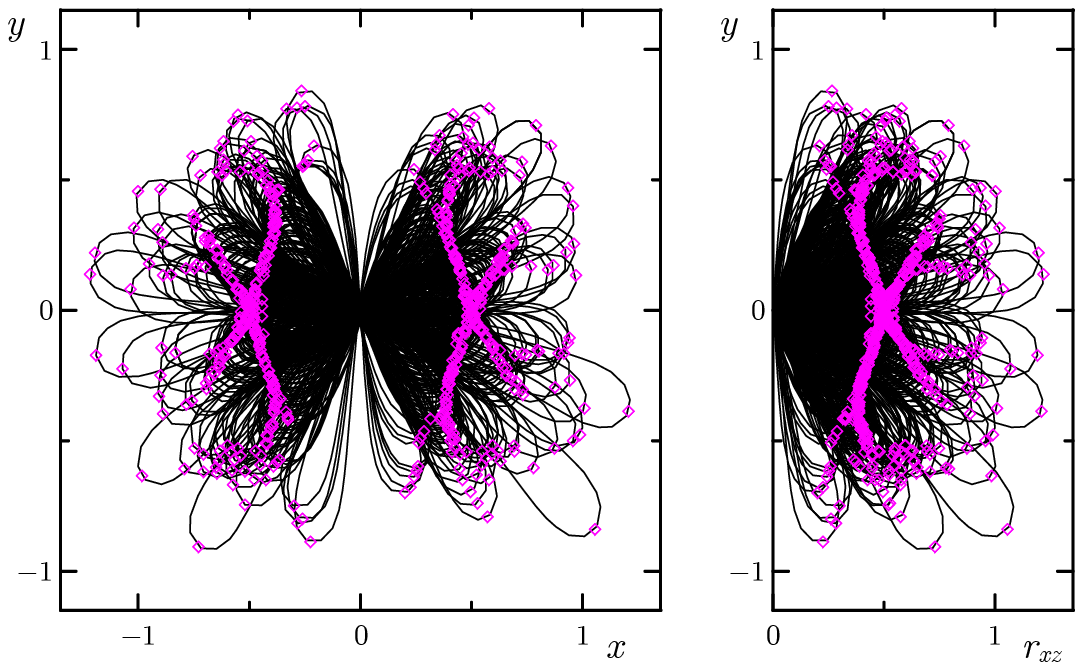}

\caption{\label{fig1}
{\it 2D setup} --- Sample trajectories (black lines) and Poincar\'e section points (magenta symbols) of overactive Brownian particle~(\ref{eq005},\ref{eq006}) (OAP) for different initial conditions in potential well (\ref{eq007}) for $U_0=1$, $V=0.05$, $\beta_2/\beta_1=\sqrt{5}-1$; $r_{xz}=|x|$. The Poincar\'e surface is given by the conditions $(\mathbf{n}\cdot\mathbf{f})=0$ and $\frac{\mathrm{d}}{\mathrm{d}t}(\mathbf{n}\cdot\mathbf{f})>0$.
%, which correspond to the local maximum of potential energy.
(a,b):~quasiperiodic trapped orbit, (c):~chaotic escape trajectory.}
\end{figure}

\begin{figure}[!t]
{\sf (a)}\hspace{-15pt}
\includegraphics[width=0.37\textwidth]{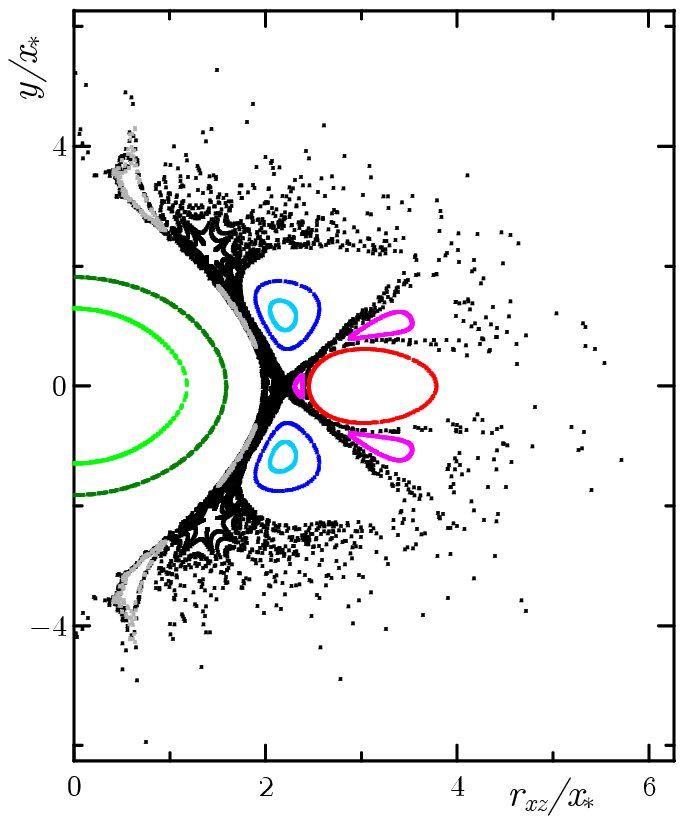}

\vspace{5pt}
{\sf (b)}\hspace{-15pt}
\includegraphics[width=0.37\textwidth]{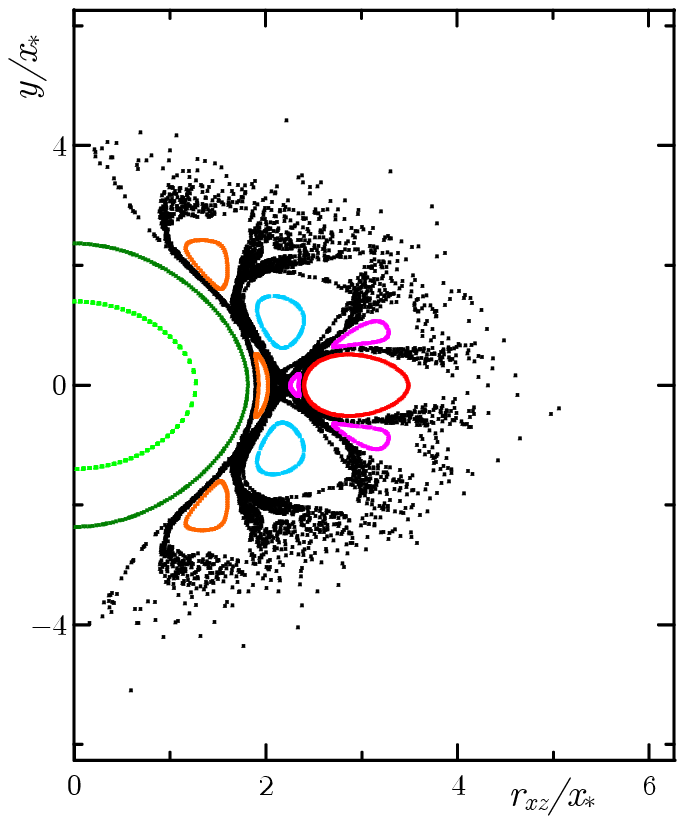}

\caption{\label{fig2}
{\it 2D setup} --- Poincar\'e sections for OAPs with different initial conditions in potential well (\ref{eq007}); colored points: quasiperiodic trapped orbits, black points: chaotic trajectories escaping to infinity. Parameters: $\beta_2/\beta_1=\sqrt{5}-1$, $V/U_0=0.05$~(a) and $0.001$~(b); $x_\ast=V/(\beta_1U_0)$.}
\end{figure}

In Fig.~\ref{fig1}, one can see sample quasiperiodic orbits~(a,b) and chaotic trajectory~(c) in a 2D system. The orbits are unambiguously represented by the Poincar\'e section (PS), given by conditions $(\mathbf{n}\cdot\mathbf{f})=0$ and $\frac{\mathrm{d}}{\mathrm{d}t}(\mathbf{n}\cdot\mathbf{f})>0$, which correspond to the local maximum of potential energy along the trajectory.
Different orbits and their types are easily visually distinguishable in this Poincar\'e section with coordinates $(|x|,y)$. In Fig.~\ref{fig2}, trapped quasiperiodic orbits are shown in color and the chaotic trajectory escaping the potential well is plotted in black. In Fig.~\ref{fig2}a, the orbits are plotted for $V/U_0=0.05$ and Gaussian-shape potential~(\ref{eq003}); in Fig.~\ref{fig2}b, $V/U_0=0.001$ and the shown part of the plane practically corresponds to potential~(\ref{eq:Ulim}), {\it i.e.}, it becomes universal for a deep potential $U_0$ or small $V$. Here, the chaotic trajectories run to infinity inspite of an infinitely growing quadratic potential.

%%%%%%%%%%%%%%%%%%%%%%%%%%%%%%%%%%%%%%%%%%%%%%%%%%%%%%%%%%%%%%%%%%%
%%%%%%%%%%%%%%%%%%%%%%%%%%%%%%%%%%%%%%%%%%%%%%%%%%%%%%%%%%%%%%%%%%%
%%%%%%%%%%%%%%%%%%%%%%%%%%%%%%%%%%%%%%%%%%%%%%%%%%%%%%%%%%%%%%%%%%%
%%%%%%%%%%%%%%%%%%%%%%%%%%%%%%%%%%%%%%%%%%%%%%%%%%%%%%%%%%%%%%%%%%%
%%%%%%%%%%%%%%%%%%%%%%%%%%%%%%%%%%%%%%%%%%%%%%%%%%%%%%%%%%%%%%%%%%%
%%%%%%%%%%%%%%%%%%%%%%%%%%%%%%%%%%%%%%%%%%%%%%%%%%%%%%%%%%%%%%%%%%%
\begin{figure}[!htb]
{\sf (a)}\hspace{-15pt}
\includegraphics[width=0.46\textwidth]{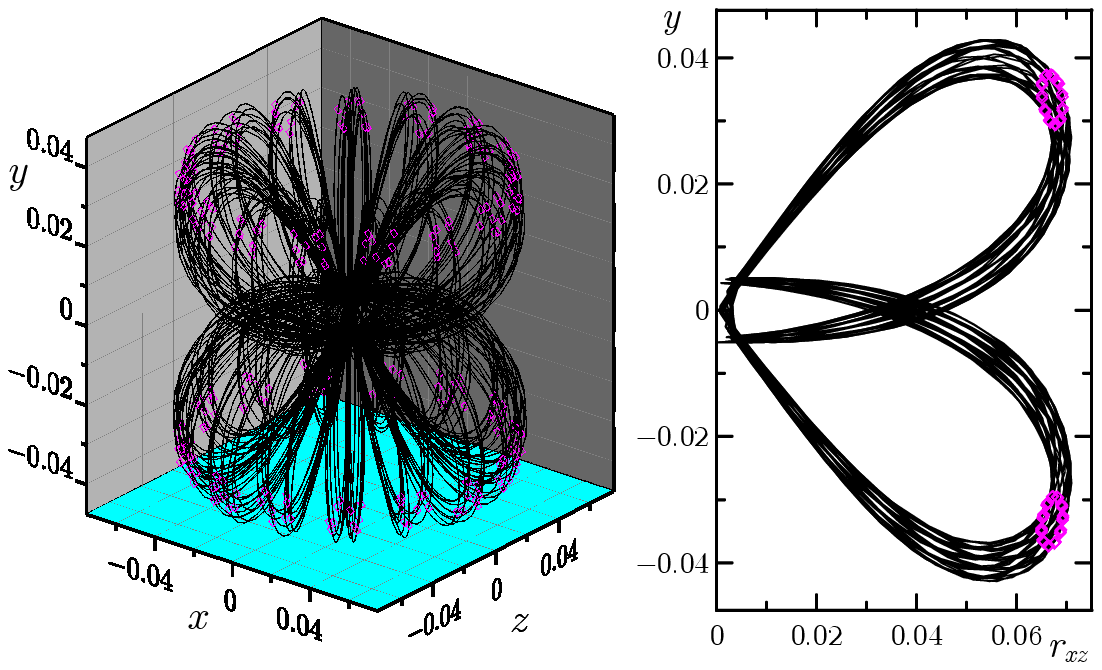}

\vspace{5pt}
{\sf (b)}\hspace{-15pt}
\includegraphics[width=0.46\textwidth]{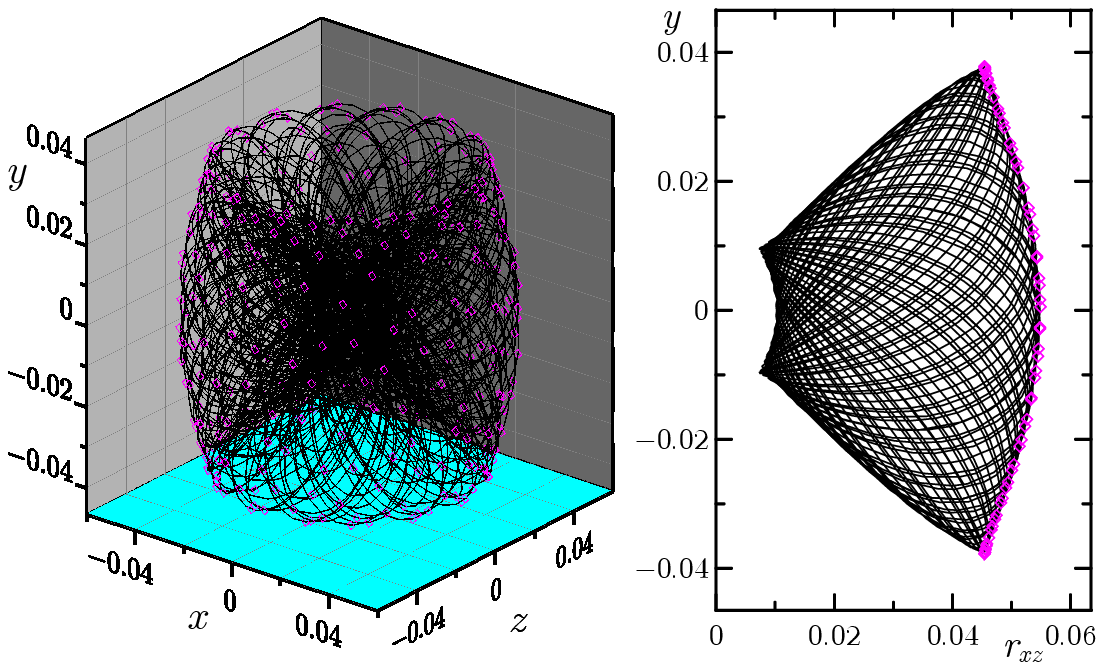}

\vspace{5pt}
{\sf (c)}\hspace{-15pt}
\includegraphics[width=0.46\textwidth]{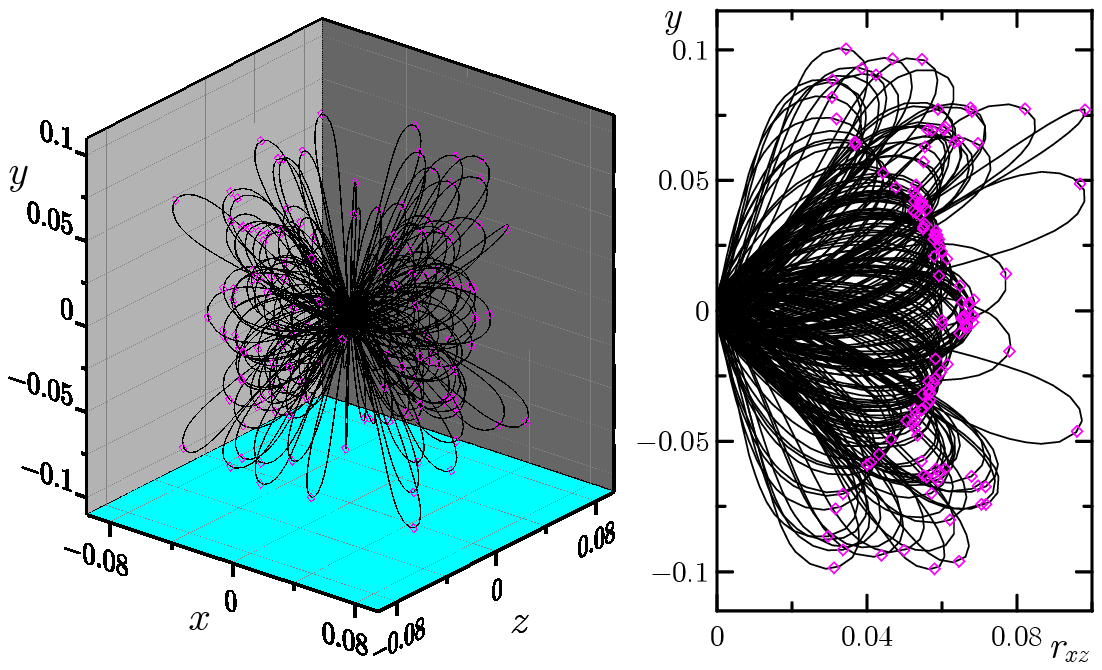}

\caption{\label{fig3}
{\it 3D setup} --- Sample trajectories (black lines) and Poincar\'e section points (magenta symbols) of OAP~(\ref{eq005},\ref{eq006}) for different initial conditions in axisymmetric potential well (\ref{eq007}) for $U_0=1$, $V=0.001$, $\beta_1=\beta_3$, $\beta_2/\beta_1=\sqrt{5}-1$, $r_{xz}=[x^2+z^2]^{1/2}$.
% The Poincar\'e surface is given by the conditions $(\mathbf{n}\cdot\mathbf{f})=0$ and $\frac{\mathrm{d}}{\mathrm{d}t}(\mathbf{n}\cdot\mathbf{f})>0$.
%, which correspond to the local maximum of potential energy.
(a,b):~quasiperiodic trapped orbit, (c):~chaotic escape trajectory.
}
\end{figure}
%%%%%%%%%%%%%%%%%%%%%%%%%%%%%%%%%%%%%%%%%%%%%%%%%%%%%%%%%%%%%%%%%%%
%%%%%%%%%%%%%%%%%%%%%%%%%%%%%%%%%%%%%%%%%%%%%%%%%%%%%%%%%%%%%%%%%%%
\begin{figure}[!htb]
{\sf (a)}\hspace{-15pt}
\includegraphics[width=0.37\textwidth]{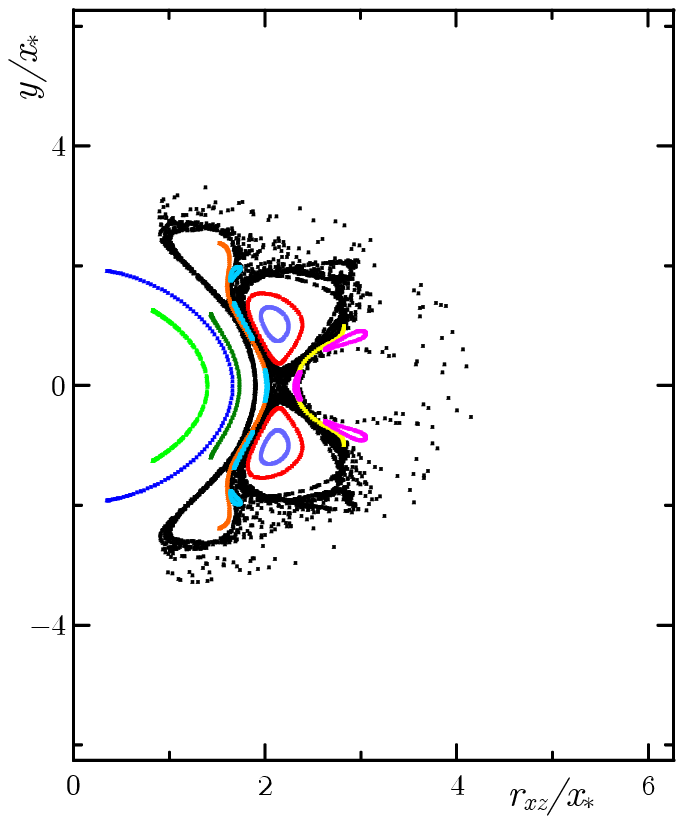}

\vspace{5pt}
{\sf (b)}\hspace{-15pt}
\includegraphics[width=0.37\textwidth]{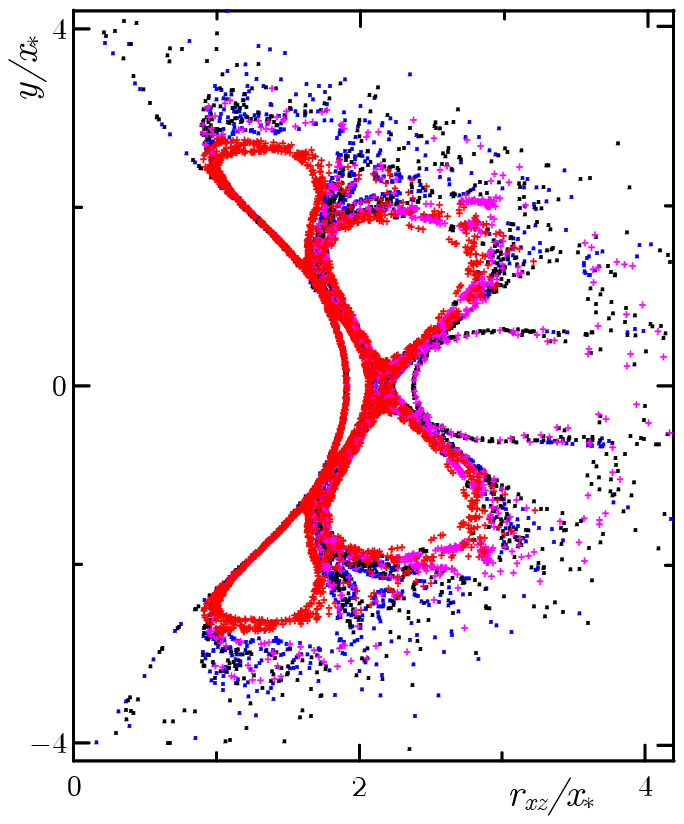}

\caption{\label{fig4}
{\it 3D setup} --- (a)~Poincar\'e sections for OAP in axisymmetric potential well (\ref{eq007}) and different initial conditions; colored points: quasiperiodic trapped state, black points: chaotic trajectories escaping to infinity. Parameters: $\beta_2/\beta_1=\sqrt{5}-1$, $\beta_3=\beta_1$, $V/U_0=0.001$; $x_\ast=V/(\beta_1U_0)$. (b)~Two sample chaotic escapes orbits for the 2D case (black and blue points) are overlapped by two sample ones (red and magenta symbols) for the axisymmetric 3D case.}
\end{figure}
%%%%%%%%%%%%%%%%%%%%%%%%%%%%%%%%%%%%%%%%%%%%%%%%%%%%%%%%%%%%%%%%%%%
%%%%%%%%%%%%%%%%%%%%%%%%%%%%%%%%%%%%%%%%%%%%%%%%%%%%%%%%%%%%%%%%%%%
\begin{figure}[!htb]
{\sf (a)}\hspace{-15pt}
\includegraphics[width=0.46\textwidth]{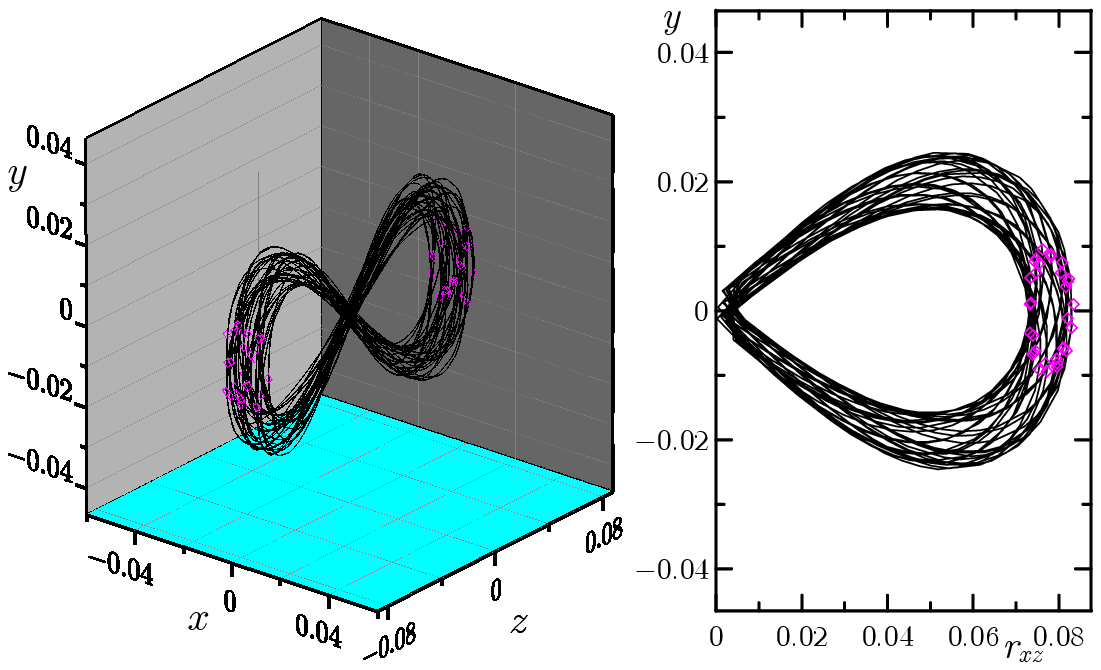}

\vspace{5pt}
{\sf (b)}\hspace{-15pt}
\includegraphics[width=0.46\textwidth]{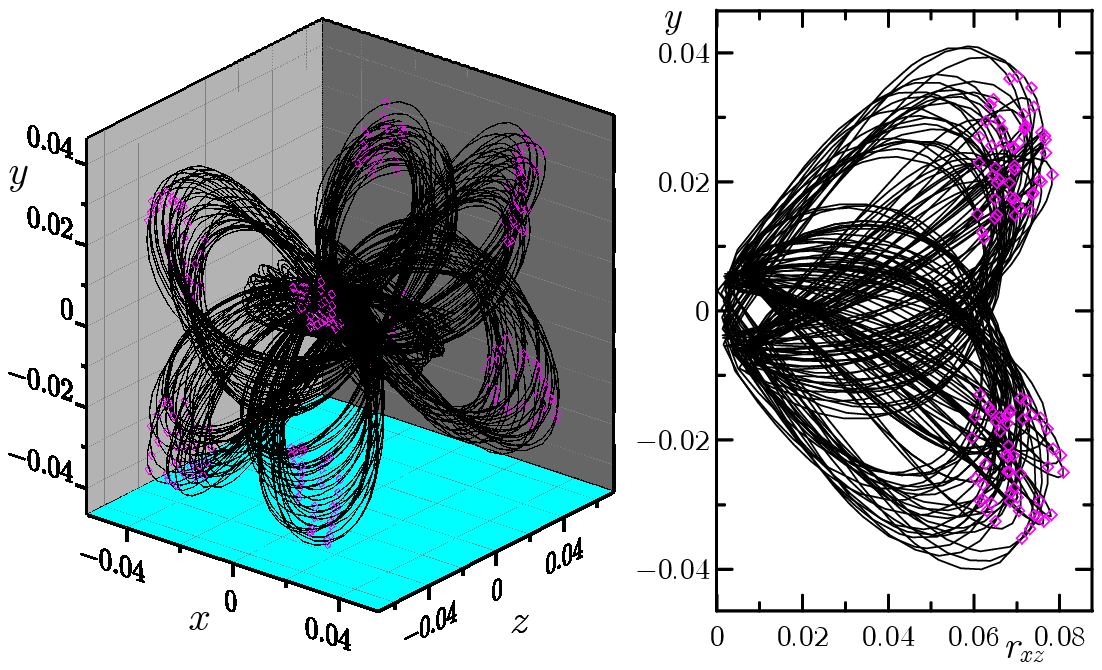}

\vspace{5pt}
{\sf (c)}\hspace{-15pt}
\includegraphics[width=0.46\textwidth]{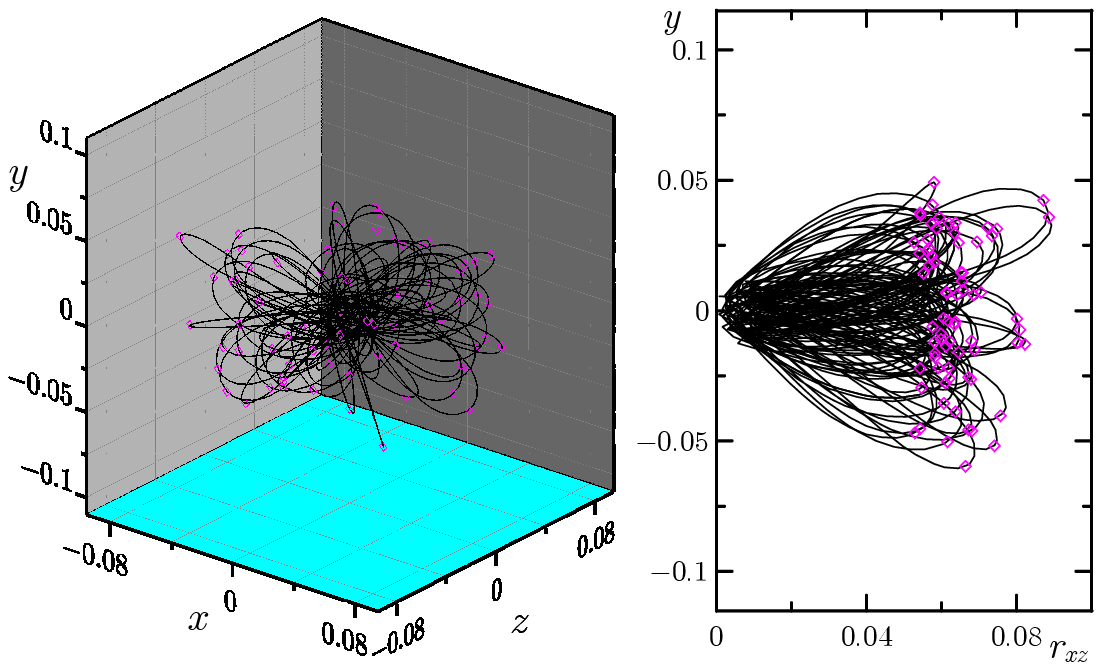}

\caption{\label{fig5}
{\it 3D setup} --- Sample trajectories (black lines) and Poincar\'e section points (magenta symbols) of OAP~(\ref{eq005},\ref{eq006}) for different initial conditions in non-axisymmetric potential well (\ref{eq007}) for $U_0=1$, $V=0.001$, $\beta_2/\beta_1=\sqrt{5}-1$, $\beta_3/\beta_1=\sqrt{3}-1$, $r_{xz}=[x^2+z^2]^{1/2}$.
% The Poincar\'e surface is given by the conditions $(\mathbf{n}\cdot\mathbf{f})=0$ and $\frac{\mathrm{d}}{\mathrm{d}t}(\mathbf{n}\cdot\mathbf{f})>0$.
%, which correspond to the local maximum of potential energy.
(a,b):~quasiperiodic orbit, (c):~chaotic trajectory.
 %Comparing the left and right panels, one can see that the quasiperiodic and chaotic trajectories can be well distinguished with the projection of the Poincar\'e section on the $(r_{xz},y)$-plane.
}
\end{figure}
%%%%%%%%%%%%%%%%%%%%%%%%%%%%%%%%%%%%%%%%%%%%%%%%%%%%%%%%%%%%%%%%%%%
%%%%%%%%%%%%%%%%%%%%%%%%%%%%%%%%%%%%%%%%%%%%%%%%%%%%%%%%%%%%%%%%%%%
\begin{figure}[!htb]
{\sf (a)}\hspace{-15pt}
\includegraphics[width=0.33\textwidth]{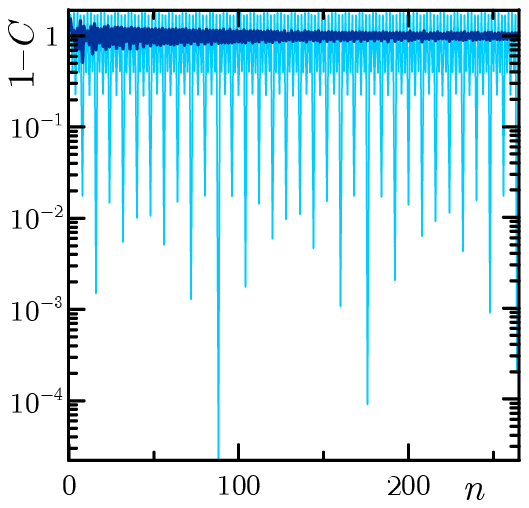}

\caption{\label{fig6}
{\it 3D setup} --- Autocorrelation functions $C(n)$ (\ref{eq:ACorr}) for the Poincar\'e map for OAP in non-axisymmetric potential well (\ref{eq007}) are plotted with the light-blue line for the quasiperiodic orbit from Fig.~\ref{fig5}(b) and with the dark-blue line for chaotic trajectory from Fig.~\ref{fig5}(c).}
\end{figure}
%%%%%%%%%%%%%%%%%%%%%%%%%%%%%%%%%%%%%%%%%%%%%%%%%%%%%%%%%%%%%%%%%%%
%%%%%%%%%%%%%%%%%%%%%%%%%%%%%%%%%%%%%%%%%%%%%%%%%%%%%%%%%%%%%%%%%%%
\begin{figure}[!htb]
{\sf (a)}\;
\includegraphics[width=0.35\textwidth]{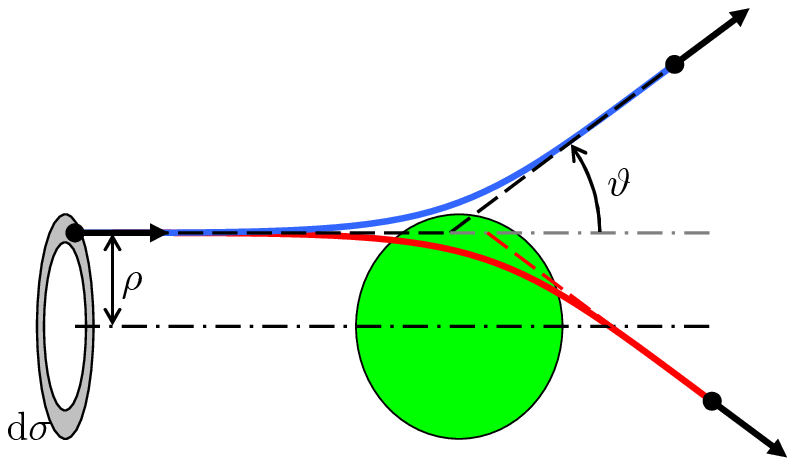}

\vspace{10pt}
{\sf (b)}\hspace{-15pt}
\includegraphics[width=0.40\textwidth]{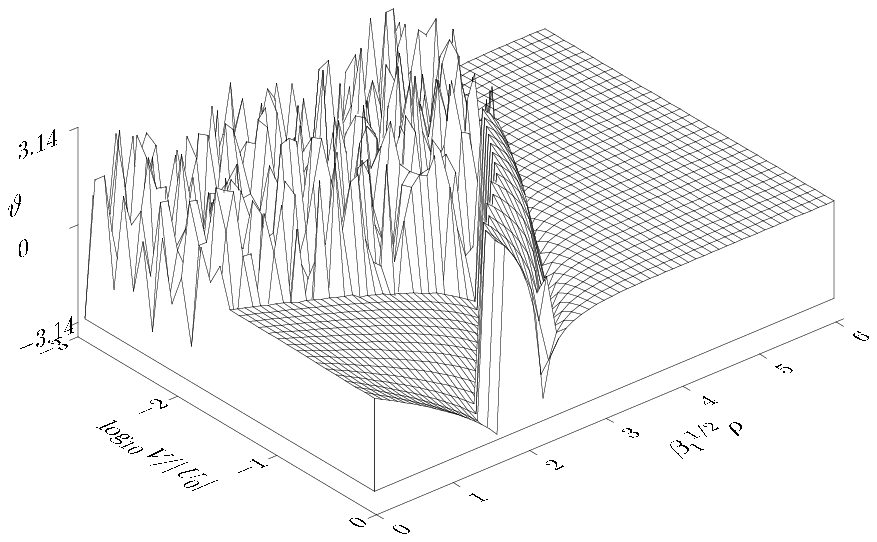}

\vspace{10pt}
{\sf (c)}\hspace{-15pt}
\includegraphics[width=0.40\textwidth]{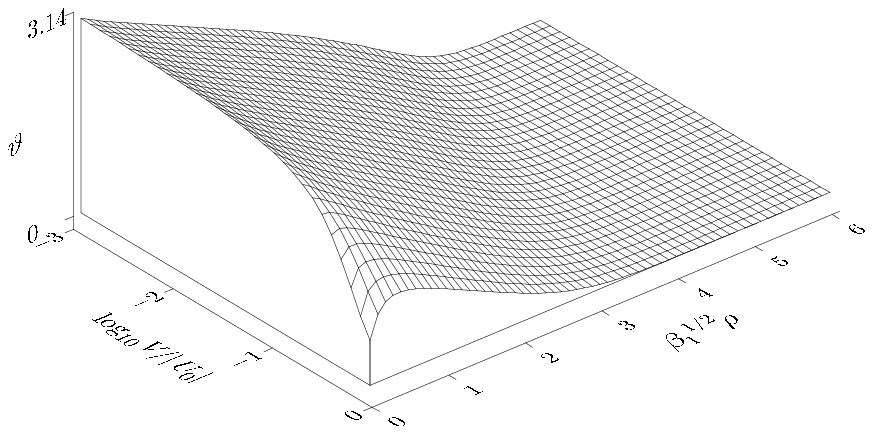}

\vspace{10pt}
{\sf (d)}\hspace{-15pt}
\includegraphics[width=0.40\textwidth]{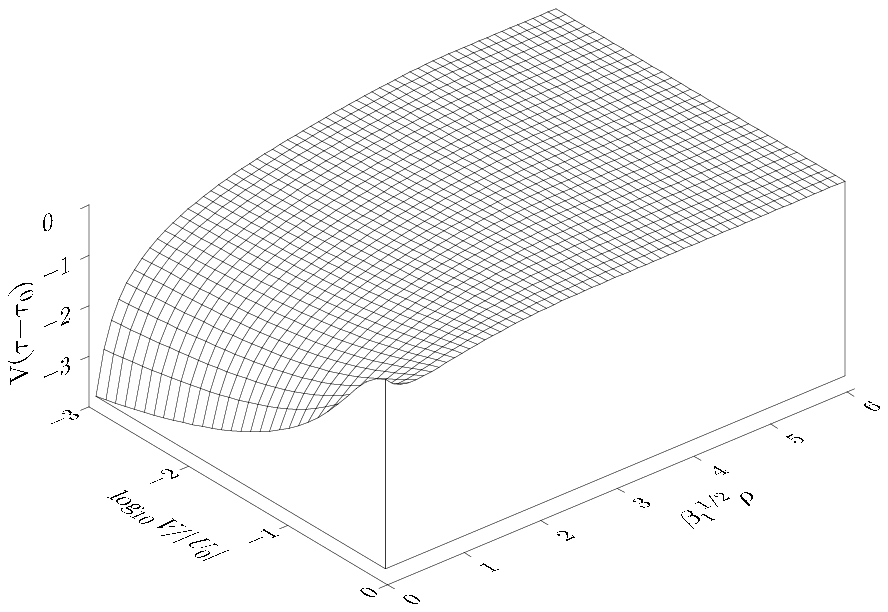}

\caption{\label{fig7}
(a)~Sketch of OAP scattering on a potential well ($U_0>0$, red trajectory) and a Gaussian obstacle (repelling potential with $U_0<0$, blue trajectory).
(b)~The direction deviation $\vartheta$ is plotted {\it versus} the self-propulsion speed $V$ and the incident parameter $\rho$ for an attractive potential~(\ref{eq003}) $U_0>0$; in the irregular area, OAP is trapped with an exponentially long residence time and chaotic trajectory. (c,d)~The direction deviation and the excess travel time $(\tau-\tau_0)$ for a repulsive potential ($U_0<0$).
}
\end{figure}
%%%%%%%%%%%%%%%%%%%%%%%%%%%%%%%%%%%%%%%%%%%%%%%%%%%%%%%%%%%%%%%%%%%
%%%%%%%%%%%%%%%%%%%%%%%%%%%%%%%%%%%%%%%%%%%%%%%%%%%%%%%%%%%%%%%%%%%

\subsection{3D axisymmetric potential}
In Fig.~\ref{fig3}, sample quasiperiodic orbits~(a,b) and chaotic trajectory~(c) are presented for axisymmetric 3D potential~(\ref{eq003}) with $\beta_2/\beta_1=\sqrt{5}-1$, $\beta_3=\beta_1$.
Comparing the left and right panels, one can see that the quasiperiodic and chaotic trajectories can be well distinguished with the projection of the Poincar\'e section on the $(r_{xz},y)$-plane.

First, noticeable changes to PS can be seen in Figs.~\ref{fig3}b and \ref{fig4}a ({\it e.g.}, see three orbits nearest to the origin): depending on the initial conditions, the section points do not reach the axis $y$. This is prevented by the motion integral~(\ref{eq:AS:1}). Indeed, as $r_{xz}\to0$, $\mathbf{f}$ becomes parallel to the $y$-axis; hence, for PS, vector $\mathbf{n}$ is perpendicular to the $y$-axis and $|M_y|=V^{-1}|\mathbf{r}_{xz}\times\mathbf{n}|e^{-U/V}=const$. The minimal value of $r_{xz}$ can be reached if the vectors in the latter product are perpendicular: $r_{xy,\mathrm{min}}=|M_y|Ve^{-U/V}$ --- PS points cannot get closer to the $y$-axis. For the initial states with smaller $|M_y|$, PS points can approach closer. This will be also important for chaotic escape trajectories, which we discuss somewhat later.

Second, PS is now defined in high-dimensional phase space, and the orbit on the 2D plane $(r_{xz},y)$ is only a projection; one can notice overlapping quasiperiodic orbit projections in Fig.~\ref{fig4}a. Overall, the images of orbits are often zero-thickness curve segments ({\it e.g.}, orange and yellow) but not a closed concentric figures (or sets of such figures) as all of them were in the 2D case (Fig.~\ref{fig2}).

Remarkably, the shape of the set of chaotic escape trajectories on the projection of PS seems identical for the 2D and axisymmetric 3D cases (see Fig.~\ref{fig4}b). However, the distribution of states prior to escape seems much more compact in the 3D case. This is expectable, as the integral of motion~(\ref{eq:AS:1}) must be zero for the trajectories with PS approaching the $y$-axis; with arbitrary initial conditions the relative likelihood to pick up such a trajectory is linearly proportional to the minimal distance $r_{xz}$ of its approach to the $y$-axis. Simultaneously, the wider-splashed chaotic trajectory segments tend to go wide also towards the $y$-axis; thus, the density of the cloud of chaotic trajectories is also diminished as one moves away from the center of the cloud of black points in Fig.~\ref{fig4}a.

\subsection{3D non-axisymmetric potential}
In Fig.~\ref{fig3}, sample quasiperiodic orbits~(a,b) and chaotic trajectory~(c) are presented for non-axisymmetric potential~(\ref{eq003}) with $\beta_2/\beta_1=\sqrt{5}-1$, $\beta_3/\beta_1=\sqrt{3}-1$. The quasiperiodic orbits still exist in this case, but become much more rare: the absolute majority of trajectories starting even from the small vicinity of the origin are chaotic. In the absence of the axial symmetry, PS is not a reliable guidance for the recognition of complicated quasiperiodic orbits; but they can be well distinguished with the autocorrelation function of the PS trajectory $\mathbf{x}(j)$:
\begin{equation}
C(n)=\frac{\langle(\mathbf{x}(j)\cdot\mathbf{x}(j+n))\rangle_j}{\langle|\mathbf{x}(j)|^2\rangle_j}\,,
\label{eq:ACorr}
\end{equation}
where $\langle\dots\rangle_j$ indicates the averaging over iteration number $j$. In Fig.~\ref{fig6}, the approach of $C(n)$ to $1$ for quasiperiodic trajectories is easily recognisable in the logarithmic scale.

\section{Scattering on potential well and obstacle}
The scattering of a particle incidenting from infinity is characterized by the scattering cross section
\[
V\mathrm{d}\sigma=2\pi\rho V\,\mathrm{d}\rho\,,
\]
where $\mathrm{d}\sigma$ is an infinitesimal cross section area, $\rho$ is the incident parameter (see Fig.~\ref{fig7}a). For a central symmetric potentials we consider in this section, the outcome of the scattering is fully given by $\vartheta(\rho)$.

In Fig.~\ref{fig7}b, the scattering diagram $\vartheta(\rho)$ is given as a function of self-propulsion speed for the attractive potential~(\ref{eq003}) with depth $U_0$. For large $\rho$, the particle passes by unaffected by the well. For moderately large $\rho$, the particle is slightly deflected towards the center of the potential well, $\vartheta<0$. One can see non-monotonous variation of $\vartheta$ as $\rho$ decreases with, obviously, formation of a loop of trajectory, which can be identified by the passage of $\vartheta$ from $\pm\pi$ to $\mp\pi$ with a continuous derivative $\partial\vartheta/\partial\rho$.

Interestingly, below certain critical value of $V$, the scattering becomes chaotic in a broad range of $\rho$, with exponentially large trapping times. For smaller $V$, the basin of attraction of the chaotic trapping regime slightly extends to bigger values of $\rho$.

In Fig.~\ref{fig7}c, the scattering diagram $\vartheta(\rho)$ is given as a function of the self-propulsion speed for the repulsive potential~(\ref{eq003}) with $U_0<0$ (obstacle). Here, the behavior is regular for all $V$. The repulsion decreases the effective passage time; $(\tau-\tau_0)$ is negative, where $\tau$ is the travel time of the particle and $\tau_0$ is the travel time with speed $V$ along the straight dashed lines (see Fig.~\ref{fig7}a).

\section{Conclusion}
The deterministic dynamics of an overactive Brownian particle under potential forces is reported. The dynamics is Hamiltonian but it is not parameterised by the value of Hamiltonian $H$; only the trajectories with $H=0$ are physically relevant. For central and axisymmetric potentials additional integrals of motion related to the rotational symmetry appear.

Quasiperiodic trapped orbits and chaotic escape trajectories are studied in 2D, axisymmetric and non-axisymmetric 3D potentials. A universal behavior for the limit of $V/U_0\to0$ (small self-propulsion speed or deep potential well) is identified. The sets of chaotic trajectories in the 2D and axisymmetric 3D cases are found to be identical; but the distribution of the probability density over these sets is profoundly different for these cases. The rotational symmetry integral of motion results in a suppression of the relative likelihood to fall on the specific trajectory proportionally to the minimal distance of its approach to the symmetry axis. As a result, the distribution of the probability density in the axisymmetric 3D case is much more compact as compared to the 2D case. In non-axisymmetric 3D case, the quasiperiodic orbits become much more rare, but still exist and the likelihood of a quasiperiodic orbit for arbitrary initial conditions is a finite value.

Scattering of the overactive Brownian particles on spherically symmetric potential wells and obstacles is characterized as a function of self-propulsion speed $V$.

\acknowledgments{The authors are thankful to A.~Pikovsky for
fruitful discussions and comments and acknowledge the financial
support from RSF (Grant No.\ 23-12-00180).
}

%%%%%%%%%%%%%%%%%%%%%%%%%%%%%%%%%%%%%%%%%%%%%%%%%%%%%%%%%%%%%%%%%%%%%%%
\begin{table}[t]
\caption{Numerical simulation performance: normalized number of mathematical operations per the dimensionless unit time time (CPU weight) {\it versus} the number $N$ of calculated derivatives for dynamical system~(\ref{eq005})--(\ref{eq007}) with $U_0=1$, $\beta_1=\beta_3=1$, $\beta_2=\sqrt{5}-1$, $V=0.0025$.}
\begin{center}
\begin{tabular}{cc}
\hline\hline
 \quad $N$ \qquad
 & CPU weight \\
\hline
12 & 39474 \\
15 & 34635 \\
16 & 34088 \\
17 & 33815 \\
18 & 33855 \\
19 & 34248 \\
20 & 34926
\\
\hline\hline
\end{tabular}
\end{center}
\label{tab1}
\end{table}
%%%%%%%%%%%%%%%%%%%%%%%%%%%%%%%%%%%%%%%%%%%%%%%%%%%%%%%%%%%%%%%%%%%%%%%

\appendix
\subsection*{Appendix: High-performance accurate numerical integration of dynamical system~(\ref{eq005})--(\ref{eq007})}
The auxiliary calculations for the Taylor coefficients, $y^{(m)}\equiv\frac{1}{m!}\frac{\partial^my}{\partial t^m}$ [notice, $(xy)^{(m)}=\sum_{l=0}^{m}x^{(m-l)}y^{(l)}$, i.e.\ no binomial coefficients unlike for the $m$th derivative of the product $(xy)$], read:

$U=-U_0\exp(-\sum_j\beta_j x_j^2/2)$\;,

$\partial{U}= -\sum_j\beta_jx_j\partial{x}_jU =-\sum_j\partial{x}_jf_j$\;,
\\
where

$f_j=\beta_jx_jU$\;;

%{\color{blue}
$U^{(m)}=-\frac{1}{m}\sum_{l=0}^{m-1}\sum_j (m-l)x_j^{(m-l)}f_j^{(l)}$
%}
\;;

%{\color{blue}
$f_j^{(m)}=\sum_{l=0}^{m}\beta_jx_j^{(m-l)}U^{(l)}$
%}
\;;

$w:=\sum_jf_jn_j$\;,

%{\color{blue}
$w^{(m)}=\sum_{l=0}^{m}\sum_j f_j^{(m-l)}n_j^{(l)}$\;;

Director  $n_j^{(m+1)}=\frac{1}{m+1}(f_j^{(m)}-\sum_{l=0}^{m}w^{(l)}n_j^{(m-l)})$\;.

Coordinate  $x_j^{(m+1)}=\frac{1}{m+1}n_j^{(m)}$\;.
%}

For the system state on the new timestep,
\[
\mathbf{x}(t+h)=\sum_{m=0}^{N}\mathbf{x}^{(m)}(t)h^m\;,
\]
\[
\mathbf{n}(t+h)=\sum_{m=0}^{N}\mathbf{n}^{(m)}(t)h^m\;,
\]
\[
\mathbf{n}(t+h):=\frac{\mathbf{n}(t+h)}{|\mathbf{n}(t+h)|}\;.
\]
Self-tuning of timestep $h$:
\[
(|\mathbf{x}^{(N)}|+|\mathbf{n}^{(N)}|)h^N<\mathrm{err}
\]
For simulations we take $\mathrm{err}=10^{-15}$. On the basis of testing (see Tab.~\ref{tab1}), we choose $N=15$, which is reasonable for calculation performance; the error per one timestep $(\mathrm{err})$ gives the error rate $(\mathrm{err}/h)\sim 10^{-13}$.

\end{document}